\documentclass[12pt]{amsart}
\usepackage{times}
\usepackage{amssymb, amsmath}
\usepackage{amsthm}
\usepackage{tikz}
\usepackage{bm}
\usetikzlibrary{matrix,arrows}
\usepackage{enumitem}
\usepackage{booktabs}
\usepackage[section]{placeins}

\theoremstyle{plain}

\theoremstyle{remark}

\graphicspath{{./images/}}

\usepackage[a4paper, total={6in, 8in}]{geometry}

\begin{document}

\title[Forecasting with Deep Learning: S\&P 500 index]{Forecasting with Deep Learning: S\&P 500 index}
\author[F. Kamalov, L. Smail, I. Gurrib]{Firuz Kamalov$^1$$^{\boldsymbol{*}}$, Linda Smail$^2$, Ikhlaas Gurrib$^3$}

\address{$^{1}$ Canadian University Dubai, Dubai, UAE.}
\email{\textcolor[rgb]{0.00,0.00,0.84}{firuz@cud.ac.ae}}

\address{$^{2}$ Zayed University, Dubai, UAE}
\email{\textcolor[rgb]{0.00,0.00,0.84}{linda.smail@zu.ac.ae}}

\address{$^{1}$ Canadian University Dubai, Dubai, UAE.}
\email{\textcolor[rgb]{0.00,0.00,0.84}{ikhlaas@cud.ac.ae}}

\date{\today
\newline \indent $^{\boldsymbol{*}}$ Corresponding author
\newline \indent DOI: 10.1109/ISCID51228.2020.00102}

\begin{abstract}
Stock price prediction has been the focus of a large amount of research but an acceptable solution has so far escaped academics. Recent advances in deep learning have motivated researchers to apply neural networks to stock prediction.
In this paper, we propose a convolution-based neural network model for predicting the future value of the S\&P 500 index. The proposed model is capable of predicting the next-day direction of the index based on the previous values of the index. Experiments show that our model outperforms a number of benchmarks achieving an accuracy rate of over 55\%.
\end{abstract}

\maketitle

\section{Introduction}
Stock price prediction is a classical problem in econometrics. There exists a plethora of methods attempting to forecast the future value of a stock. However, according to the predominant economic theory - the Efficient Market Hypothesis -  all the publicly available information is already incorporated in the current stock price. The theory postulates that the future stock price is  practically impossible to predict. Despite the gloomy outlook, we are able to construct a deep learning model that predicts the direction of the S\&P 500 index with a nontrivial rate of accuracy.

The new surge of computational power, and the availability of mass data from financial markets has generated a huge interest in the potential use of Machine Learning (ML) in improving market forecasting when compared with traditional approaches \cite{Strader, Cavalcante}.  Artificial neural networks, support vector machine, genetic algorithms combined with other techniques have been used for the past decade to forecast and analyze financial markets because these techniques do not require any assumptions about the data and often achieve higher accuracy than traditional methods.   In 2009, the authors in \cite{Atsalakis} surveyed more than 100 articles and concluded that neural networks (NNs) improve market forecasting, when compared to traditional approaches.  Many other researchers used ML models such as NNs and support vector regression machines (SVR) along with feature selection algorithms to select the input variables, while providing empirical evidence that feature selection algorithms improved their  model’s performance \cite{Lee, Barra, Papadimitriou} .  Lately, deep learning has increasingly become a popular tool to forecast stock markets due to its good nonlinear approximation capability and adaptive self-learning \cite{Zhi, Shen} . Artificial neural networks (ANNs) using different deep learning algorithms are classified as deep neural networks (DNNs) and have been used in many important domains including financial markets \cite{Zhong}. 

We propose a network containing two hidden layers: convolutional and fully connected. The model uses the previous closing values and volume of the index to predict the next-day direction. The key insight in our approach is the utilization of convolutional layer which facilitates consideration of individual instances in time series in the context of their  temporal  neighbors. The application of convolutional layer in time series data proves to be an effective technique. To test our model we compare it against 7 benchmark models. The results of the experiments demonstrate that the proposed model outperforms the benchmark models both in accuracy and precision. We believe the proposed neural network architecture would be a useful tool for researchers interested in time series forecasting. 

Our paper is organized as follows. In Section 2, we provide a brief discussion of the current literature. In Section 3, we present the proposed network architecture. Section 4 contains the results of the numerical experiments. And Section 5 concludes the paper.

\section{Literature}
Stock price forecasting is a classic problem. There exists a plethora of literature devoted to the subject.
To predict the trends of the US market convolutional neural networks models such as, AlexNet, ResNet, and Inception were used in \cite{Barra}. Those models were fed using a multi-resolution Gramian angular fields images, generated from time series related to the Standard and Poor's 500 index future. The results showed that this method outperformed the buy-and-hold strategy in a time frame where the latter provided excellent returns.  
The severe changes of the S\&P 500 index could be either extreme positive or negative returns of the index are usually called spikes. Using 1860 daily observations collected over 9 years period, the authors in \cite{Papadimitriou} forecast the spikes of the S\&P 500 stock returns using Support Vector Machines methodology (SVM). The results showed that the optimum predictive SVM model was the one using the RBF kernel along with two lags of the S\&P 500 returns. The model achieved around 71\% forecasting accuracy for the spikes .
To predict the future price of the stock index, a NN was proposed in \cite{Lee}  where the authors used only the data of individual companies to obtain enough data for training the NN instead of using the target stock index data. Then the trained model and a sliding window technique were used to forecast short-term stock values. The process considered a heuristic to control the possible extrapolation anomalies of the DNN. The results showed that the NN trained in this manner performed better than the NNs trained on the S\&P 500 stock index data.  
In \cite{Zhong}, the authors predicted the daily revenue of a set of stocks based on a 10-year collected data about the SPDR S\&P 500 ETF, along with 60 financial indicators using ANNs and DNNs classifiers over the entire untransformed dataset and the principal component analysis PCA-represented datasets.  The results of this dimensionality reduction study showed that the DNNs using two PCA-represented datasets gave higher accuracy than those using the entire untransformed dataset, as well as several other hybrid machine learning algorithms. In fact, the accuracy of the DNNs was demonstrated to be higher as the number of the hidden layers increased from 12 to 1000.  

Stock returns have also been used in directional price movement prediction.
In \cite{Fischer}, the authors used LSTM to predict directional movements for the constituent stocks of the S\&P 500 market index.  The authors use sequences of one-day returns as inputs for the LSTM model. The length of each input sequence  is 240 corresponding to the daily returns over 240 days prior to the forecast date. The proposed method is found to outperform random forest, deep neural net, and logistic regression models. 
In \cite{Liew}, the authors study the effectiveness of three feature subsets - returns, volume, and days - on the performance of directional forecasting models. The analysis is done on a 5-year ETF data using DNNs, RFs, and SVMs. The authors discover that volume is an important factor in forecasting. In \cite{Kamalov3}, the authors applied daily stock returns to forecast significant price changes using various neural network architectures. The results indicate that LSTM models can be effective in predicting significant price changes based on the prior temporal data.

One of the issues that arise when analyzing stock data is uneven distribution of price changes. If the stock data is obtained from a period of economic growth than the number of days with positive price change would outnumber the number of days with negative price change. The skewed distribution of price changes can have a negative effect on the performance of the classification algorithms \cite{Thabtah}. A common approach to address class imbalance is to balance the data through resampling \cite{Kamalov1}. Alternatively, outlier detection methods can be used in place of the standard classification algorithms \cite{Kamalov2}.

\section{Model specifications}
In this section, we present the details of the proposed network architecture for predicting the direction of the S\&P 500 index. We employ the daily closing values and trading volume from the previous 14 days to forecast the next-day direction of the index. The model contains two hidden layers: convolutional and fully connected. Convolution operation takes a weighted sum of the points in the neighborhood of the center point. As a result, each point in the sequence is considered in the context of its neighboring points. 
\subsection{Model architecture}
\begin{figure}[h!]
\centering
\includegraphics[width=0.7\textwidth]{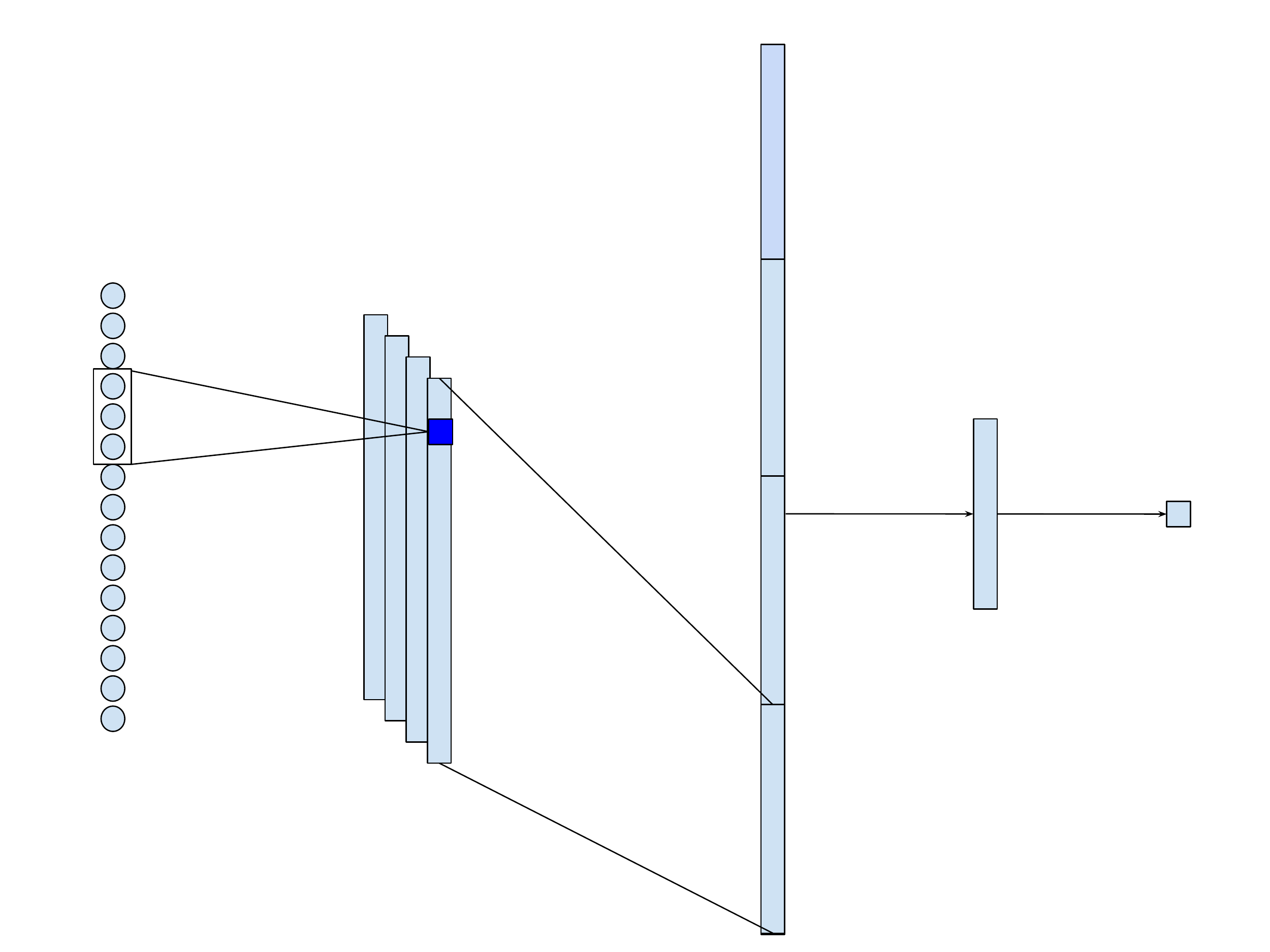}
\caption{The architecture of the proposed neural network model for predicting the S\&P 500 index.}
\label{conv1fc}
\end{figure}

The details of the proposed architecture are presented in Figure \ref{conv1fc}. As shown in the figure, the model consists of a 14-dimensional input layer followed by a convolutional layer, a flattening layer, a fully connected layer, and a 1-dimensional output layer. The input layer represents the closing values and volume from the previous 14 trading days. The convolutional layer consists of four filters of size 3. As a result, the data for each day in the input layer is considered in the context of its preceding and following days. This technique allows  to obtain more informative features in the model. Finally, we utilize a fully connected layer to analyze the output of the convolutional layer and make a prediction. Rectified linear unit (ReLU) is employed as the activation function in every layer of the network.

\subsection{Model training}
The model is trained on daily index values from the past 30 years. The data is divided into train/validation/test sets according to 70/15/15 ratio. We apply the RMSprop optimizer with batch size 32 to train our model. The RMSprop algorithm takes into account the local topology of the cost function and adjust the gradient descent process correspondingly. The batch size of 32 is chosen to allow the optimization algorithm to explore a wider range of values in search of the minima. We apply early stopping to avoid model overfitting. Concretely, the validation error is monitored during the training and the process is stopped when the validation error stops decreasing for more than 10 epochs. The shallow architecture of the network serves as another regularization tool - with fewer number of parameters the model is less likely to overfit. The model is implemented using Keras API on Colab platform. 
\section{Numerical experiments}
In this section, we present the results of numerical experiments that were carried out to test the efficacy of the proposed model. We benchmark the proposed model against 7 other network architectures. The results demonstrate that the proposed model outperforms the benchmark models achieving the highest accuracy rate in predicting the next-day direction of the index. 
\subsection{Implementation}
The data used in our experiment covers the period of 1990-07-15 to 2020-07-15. The data was obtained from the Yahoo Finance. The model inputs are the closing and volume values from the previous 14 trading days. Upon reshaping the data size is (7545, 14, 2), where 7545 is the number of days, 14 is the length of each input sequence, and 2 is the number of features. The model is trained in 2 stages. First, the model is trained on the training set using early stopping callback. The training is stopped once the validation error ceases to decrease. Using the number of epochs obtained from early stopping we retrain the model using the combined train and validation sets. After retraining the model we evaluate it on the holdout test set.  
\subsection{Benchmark models}
\begin{figure}[h!]
\centering
\includegraphics[width=0.7\textwidth]{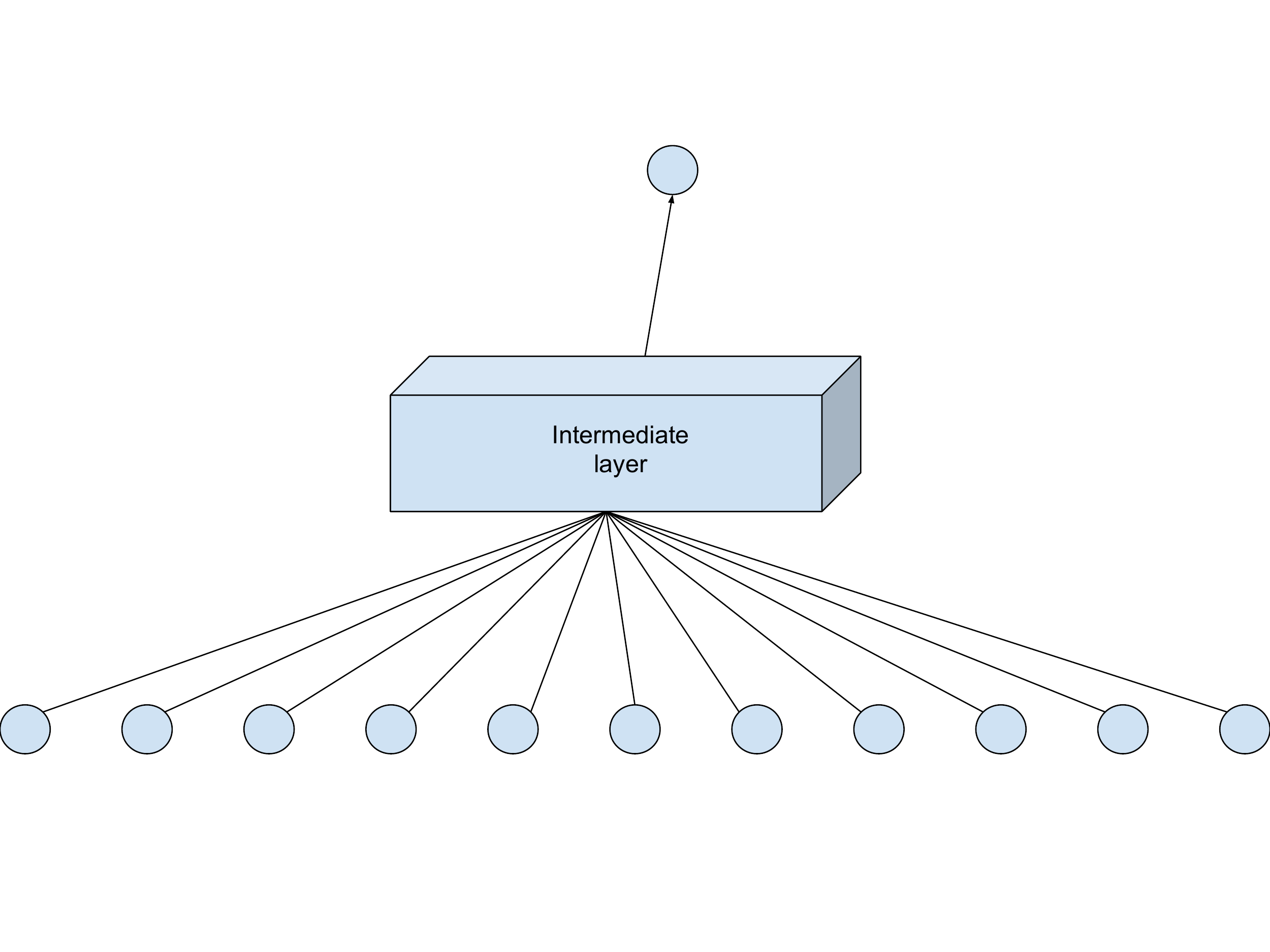}
\caption{The general structure of the benchmark neural networks used in the experiments.}
\label{nn_diagram2}
\end{figure}

We employ 7 benchmark models to compare against the proposed model. The list of models includes a variety of different architectures such as fully connected, recurrent, and convolutional layers. The general structure of the benchmark models follows that of the proposed model (Figure~\ref{nn_diagram2}). As shown in Figure~\ref{nn_diagram2}, each model consists of a 14-dimensional input layer and a 1-dimensional output layer. The input layer represents the values of the index from the previous 14 trading days while the output layer represents the predicted index value on the next day. 

We employ 7 different benchmark architectures for the intermediate layer in Figure~\ref{nn_diagram2}.  The details of the models are presented in Table \ref{tab:models}. In the simplest case, we only use fully connected layers. A single fully connected layer is flexible enough - given enough neurons - to approximate any real valued continuous function on a compact set. However, the underlying assumption when using fully connected layers is that the order of the inputs is irrelevant. Since we are dealing with a time series the order does matter. Hence, we employ recurrent neurons. Recurrent neurons take into account the sequential nature of time series data. As a result, RNN and its variant LSTM are popular architectures in time series analysis.

\begin{table}[h!]
\centering
\caption{The descriptions of the benchmark models follow Keras nomenclature.}
\label{tab:models}
\begin{tabular}{lll}
\toprule
index & {model} &  description  \\
\midrule
1 & \texttt{fc1} &     Dense(14) \\
2 & \texttt{fc2} &    Dense(14) $\rightarrow$ Dense(7) \\
3 & \texttt{rnn1} &     RNN(4)  \\
4 & \texttt{rnn1fc} &    RNN(4) $\rightarrow$ Dense(4) \\
5 & \texttt{rnn2} &     RNN(6)  \\
6 & \texttt{lstm1} &    LSTM(6) \\
7 & \texttt{conv1} &    Conv1D(4, 3) \\
\bottomrule
\end{tabular}
\end{table}
\subsection{Results}
\begin{figure}[h!]
\centering
\includegraphics[width=0.7\textwidth]{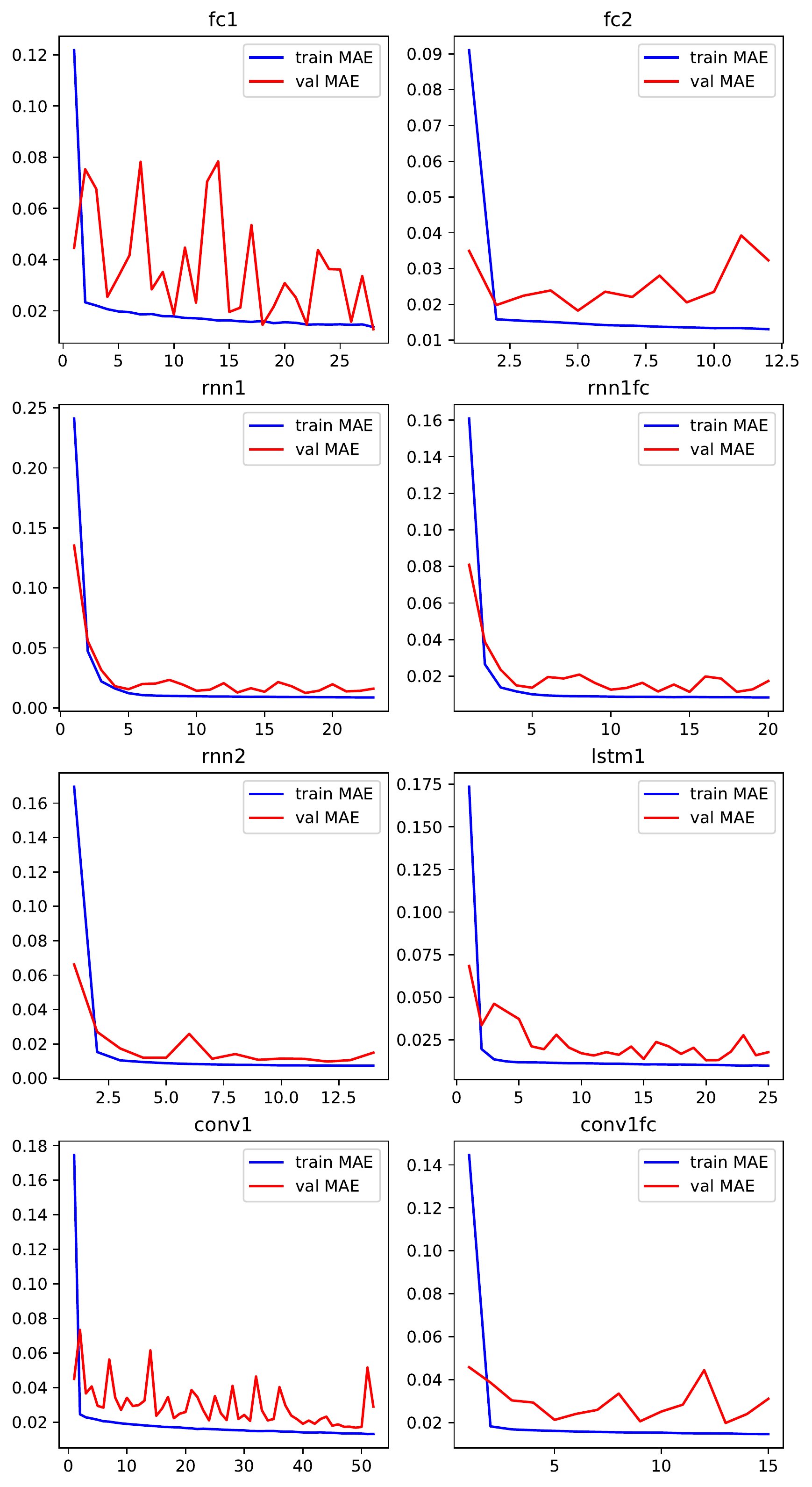}
\caption{Training results of the proposed and benchmark models. The proposed model is denoted by \texttt{conv1fc}. The horizontal axis shows the number of training epochs.}
\label{rms32}
\end{figure}

The training results are presented in Figure \ref{rms32}. We observe that while the training error quickly reaches its near minimum the validation error meanders up and down near the training curve. Almost all models, with exception of \texttt{conv1}, stopped training after 15-25 epochs. As mentioned above, the training stops once the validation error ceases to decrease. The proposed model, denoted by \texttt{conv1fc}, achieves the early stopping criterion in just 15 epochs.

The accuracy results of predicting the next-day direction of the S\&P 500 index are presented in Table \ref{tab:dir}. Note that the proposed model achieves the top accuracy rate of 56.21\%. It is 6\% higher than the average accuracy rate of the benchmark models and 1\% higher that the second best accuracy rate. Another common benchmark for predicting the direction of stock price is the naive strategy of random guessing which yields accuracy rate of 50\%. It further shows that our model accuracy rate of 56.21\% is significant.

\begin{table}[h!]
\centering
\caption{Predicting the direction of S\&P 500 index using the proposed (\texttt{conv1fc}) and benchmark models. Note that the proposed model achieves the top accuracy rate of 56.21\%.}
\label{tab:dir}
\begin{tabular}{lr}
\toprule
Model &    Accuracy  \\
\midrule
\texttt{fc1} &     0.5433 \\
\texttt{fc2} &    0.5004 \\
\texttt{rnn1} &    0.5523  \\
\texttt{rnn1fc} &    0.4602 \\
\texttt{rnn2} &      0.5103\\
\texttt{lstm1} &    0.4620 \\
\texttt{conv1} &    0.4853  \\
\texttt{conv1fc} &   \textbf{0.5621} \\
\bottomrule
\end{tabular}
\end{table}
\section{Conclusion}
In this paper, we proposed a deep learning architecture for time-series forecasting and used it to successfully predict the next-day direction of the S\&P 500 index. The key insight of our model is the use of convolutional layer consisting of 
four filters of size 3. As a result, the data for each day in the input layer is considered in the context of its preceding and following days. The application of the filters leads to more informative features in the hidden layer. The results of numerical experiments against 7 benchmark models demonstrate the efficacy of the proposed model. Our model achieves the top accuracy rate of 56.21\%. The proposed deep learning architecture can be applied to study other sequential data. We believe it would be of interest to researchers and practitioners in the field of time-series forecasting.


\begin{thebibliography}{10}

\bibitem{Atsalakis} Atsalakis, G. S., \& Valavanis, K. P. (2009). Surveying stock market forecasting techniques—Part II: Soft computing methods. Expert Systems with Applications.

\bibitem{Barra} Barra S., Carta S. M., Corriga A., Podda A. S. (2020). Recupero D. R., "Deep learning and time series-to-image encoding for financial forecasting," in IEEE/CAA Journal of Automatica Sinica, vol. 7, no. 3, pp. 683-692.

\bibitem{Cavalcante} Cavalcante R.C., Brasileiro R.C., Souza V.L.F., Nobrega J.P.,\& Oliveira A.L.I., Computational Intelligence and Financial Markets: A Survey and Future Directions. Expert Systems with Applications. 2016 Aug; 55  (15):194–211.

\bibitem{Fischer}Fischer, T., \& Krauss, C. (2018). Deep learning with long short-term memory networks for financial market predictions. European Journal of Operational Research, 270(2), 654-669.

\bibitem{Kamalov1} Kamalov, F. (2020). Kernel density estimation based sampling for imbalanced class distribution. Information Sciences, 512, 1192-1201.

\bibitem{Kamalov2} Kamalov, F., \& Leung, H. H. (2020). Outlier detection in high dimensional data. Journal of Information \& Knowledge Management, 2040013.


\bibitem{Kamalov3} Kamalov, F. (2020). Forecasting significant stock price changes using neural networks. Neural Computing and Applications. 1-13.

\bibitem{Lee} Lee J, \& Kang J (2020) Effectively training neural networks for stock index prediction: Predicting the S\&P 500 index without using its index data. PLoS ONE 15(4): e0230635.


\bibitem{Liew}Liew, J. K. S., \& Mayster, B. (2017). Forecasting etfs with machine learning algorithms. The Journal of Alternative Investments, 20(3), 58-78.


\bibitem{Papadimitriou} Papadimitriou, T., Gogas, P. \& Athanasiou, A.F. (2020). Forecasting S\&P 500 spikes: an SVM approach. Digital Finance.


\bibitem{Shen} Shen Z., Zhang Y., Lu J., Xu J., \& Xiao G. (2020). A novel time series forecasting model with deep learning, Neurocomputing. 396  302–313.

\bibitem{Strader} Strader, T. J., Rozycki, J. J., Root, T. H., \& Huang, Y. (. (2019). Machine learning stock market prediction studies: Review and research directions. Journal of International Technology and Information Management, 28(4), 63-83.

\bibitem{Thabtah}Thabtah, F., Hammoud, S., Kamalov, F., \& Gonsalves, A. (2020). Data imbalance in classification: Experimental evaluation. Information Sciences, 513, 429-441.

\bibitem{Zhi} Zhi S. U., Man L. U., \& Dexuan L. I. (2017). Deep Learning in Financial Empirical Applications: Dynamics, Contributions and Prospects. Journal of Financial Research.  

\bibitem{Zhong} Zhong, X., \& Enke, D. (2019). Predicting the daily return direction of the stock market using hybrid machine learning algorithms. Financial Innovation, 5(1), 4. 

\end{thebibliography}
\end{document}